# Size-dependent electronic-transport mechanism and sign reversal of magnetoresistance in $Nd_{0.5}Sr_{0.5}CoO_3$

S. Kundu[a] and T. K. Nath[*]

*Department of Physics and Meteorology, Indian Institute of Technology, Kharagpur, West Bengal 721302, India*
[a]*email: souravphy@gmail.com*

## Abstract

*A detailed investigation of electronic-transport properties of $Nd_{0.5}Sr_{0.5}CoO_3$ has been carried out as a function of grain size ranging from micrometer order down to an average size of 28 nm. Interestingly, we observe a size induced metal-insulator transition in the lowest grain size sample while the bulk-like sample is metallic in the whole measured temperature regime. An analysis of the temperature dependent resistivity in the metallic regime reveals that the electron-electron interaction is the dominating mechanism while other processes like electron-magnon and electron-phonon scatterings are also likely to be present. The fascinating observation of enhanced low temperature upturn and minimum in resistivity on reduction of grain size is found due to electron-electron interaction (quantum interference effect). This effect is attributed to enhanced disorder on reduction of grain size. Interestingly, we observed a cross over from positive to negative magnetoresistance in the low temperature regime as the grain size is reduced. This observed sign reversal is attributed to enhanced phase separation on decreasing the grain size of the cobaltite.*

---

[*]Corresponding author: tnath@phy.iitkgp.ernet.in

Tel: +91-3222-283862
Area code: 721302
INDIA

## 1. Introduction

Cobaltites are one of the most interesting systems amongst the strongly correlated electron materials having various fascinating electronic and magnetic properties like manganites. Similar to manganites, cobaltites posses close interplay between charge, orbital, spin and lattice degrees of freedom [1-5]. However, it has one extra degree of freedom, the spin state of Co ion. Multiple spin state found in Co ion is a consequence of comparable value of the crystal field splitting energy and Hund's coupling energy [1-5]. In a material like $La_{1-x}Sr_xCoO_3$ the $Co^{3+}$ and $Co^{4+}$ ions may have three different spin states each namely, the low spin (LS: $t_{2g}^6e_g^0$ for $Co^{3+}$ and $t_{2g}^5e_g^0$ for $Co^{4+}$), the intermediate spin (IS: $t_{2g}^5e_g^1$ for $Co^{3+}$ and $t_{2g}^4e_g^1$ for $Co^{4+}$) and the high spin (HS: $t_{2g}^4e_g^2$ for $Co^{3+}$ and $t_{2g}^3e_g^2$ for $Co^{4+}$) state [1-5]. Spin state transition may occur under influences like change of temperature [2,5]. The observed ferromagnetism in cobaltites originates most likely from double exchange interaction between $Co^{3+}$ and $Co^{4+}$ species similar to the case of doped manganites, but dependent on Co spin state. Another most interesting feature of cobaltites is the phase separation effect [4,6-8] like that of manganites. The low temperature phase of cobaltite materials like $La_{1-x}Sr_xCoO_3$ have been found to be inhomogeneous where ferromagnetic (FM) regions coexist with various non-FM regions or spin glass regions [7,8]. Though $La_{1-x}Sr_xCoO_3$ is one of the most studied cobaltite systems, other class of rare earth cobaltites (i.e. Pr, Nd, Eu etc. based systems) [9-13] have also been investigated to some extent. Like the La-based cobaltites, the system $Nd_{1-x}Sr_xCoO_3$ with a lower bandwidth is also known to have a variety of magnetic and electronic properties depending on the hole doping level. For instance, in the low doped regime this material displays a phase separated state and glassy magnetic behaviour [9,10]. This system also exhibits ferrimagnetic ordering below certain temperature in a particular doping regime [14]. As far as electronic-transport property is concerned $Nd_{1-x}Sr_xCoO_3$ display insulating behaviour in the low doped regime and metallic behaviour in heavily doped regime [9]. However, in any case, the physics of cobaltites remained less understood compared to that of manganites till date. Especially, regarding the recent trend of investigation on materials with reduced dimension (nanomaterials), cobaltites are comparatively less explored. We here present our results of investigations on different magnetic and electronic-transport properties of $Nd_{0.5}Sr_{0.5}CoO_3$ in a wide range of grain size of this material. According to earlier reports this system in bulk form is metallic up to the room temperature and undergoes a transition into a ferromagnetic state below around 225 K [9, 15]. On reduction of grain size down to 28 nm, a complete reversal of electronic- and magneto-transport properties is observed, while a moderate influence on magnetic properties is also found.

## 2. Experimental details

We have synthesized our polycrystalline $Nd_{0.5}Sr_{0.5}CoO_3$ samples with varying grain size through chemical route known as pyrophoric reaction [16]. We have employed high purity $Nd_2O_3$, $Sr(NO_3)_2$ and $Co(NO_3)_2.6H_2O$ in stoichiometric proportion and dissolved in distilled water with proper amount of $HNO_3$. Triethnolamine (TEA) was then added to the solution with a molar ratio to the metal ions = 4:1:1 (Nd/Sr:Co:TEA=1:1:4). This solution was stirred and heated at 180 $^0$C until combustion took place. The obtained fluppy powder was divided in four parts, ground thoroughly and calcinated in air (5 hours) at different temperatures to obtain polycrystalline $Nd_{0.5}Sr_{0.5}CoO_3$ samples with different grain size. The powder samples were pressed into pellets and again sintered in air at the same temperature for 1 hour. The samples are designated as S1150, S950, S850 and S750 after their corresponding sintering temperature in degree Celsius. The phase of the samples was checked by high resolution x-ray diffraction technique. The average grain size was estimated by employing transmission electron microscope (TEM) and field emission scanning electron microscope (FESEM). The electronic-transport measurements were carried out employing a closed cycle helium refrigeration cryostat fitted inside a superconducting magnet (8 T) in the temperature range of 2-300 K (Cryogenics Ltd., U.K.). The magnetic properties of the samples have been investigated employing homemade vibrating sample magnetometer (77-300 K) [17] and homemade ac susceptometer (77-300 K). Both these instruments employ high precision Lock-in-amplifier (SR830) for signal detection and PID temperature controller (Lakeshore 331S) to control and monitor the sample temperature.

## 3. Results and discussion

The room temperature x-ray diffraction patterns of S1150 and S750 samples are shown in the Fig. 1. No impurity phase is observed in the x-ray diffraction patterns of these two as well the other intermediate grain size samples. The x-ray data of all the samples were refined using Rietveld method assuming *Imma* space group of orthorhombic structure. The fitted curves are also shown in the Fig. 1. The obtained lattice parameters and unit cell volumes are listed in Table1. We observe that the change in lattice parameters and unit cell volume is very small with the change of grain size. This change is also non-monotonic with grain size. This indicates that the grain size have only negligible influence on structural distortions. The TEM image of S750 sample shown in the Fig. 2 (a) displays nanometric size of the grains. The distribution of grain size plotted in the inset of Fig. 2(a) displays log normal type distribution. The average grain size of S750 is found to be about 28 nm. The high resolution lattice image of S750 in the Fig. 2 (b) exhibits that the crystallographic structure is not prominent at the boundary region of the grains (marked by arrows) indicating the disorder in the grain surfaces. The FESEM images of S850, S950 and S1150 samples are shown in the Fig. 2 (c), (d) and (e), respectively. The average grain sizes estimated from the plot of grain size distribution of these samples are given in the Table1. The S1150 sample has average grain size in the micrometer regime. This indicates that the sample should display bulk-like behaviour. Further, we have experimentally checked the stoichiometry of our samples through EDS (Energy-dispersive x-ray spectroscopy) measurements. There is a possibility of change in stoichiometry in the samples due to the change in calcination temperature. The EDS curves of S1150 and S750 samples are shown in the Fig. 3. From these spectra the calculated chemical formula of the samples are found to be $Nd_{0.51}Sr_{0.47}Co_{1.02}O_{2.99}$ for S1150 and $Nd_{0.51}Sr_{0.48}Co_{1.01}O_{2.99}$ for S750 samples, respectively. We find that there is negligibly small change of stoichiometry in the samples when the grain size is varied by changing calcinations temperature.

The dc magnetic studies of the samples have been carried out in terms of measuring temperature dependent magnetization. The data was taken at 100 Oe field during heating after cooling the sample at zero field (ZFC) or at 100 Oe field (FC) as shown in the Fig. 4. All the samples display a FM-paramagnetic (PM) transition at around 225 K ($T_C$). This transition temperature is in agreement with previously reported value of this system [9]. Though $T_C$ is not found to change much with the variation of grain size, the S750 sample displays a much broader FM-PM transition compared to the other samples. This may be attributed to the presence of grain size distribution. One interesting observed feature is the huge bifurcation between ZFC and FC magnetization of all the samples. This indicates that the low temperature ferromagnetic phase is highly anisotropic in nature. This anisotropic magnetic behaviour in cobaltites is very common and has been discussed earlier [18,19]. This is mainly magnetocrystalline anisotropy added with the surface anisotropy originating from surface disorder. One possible origin of the magnetic anisotropy is the presence of Co ions as pointed out by different groups earlier [19]. We find that the ZFC magnetization falls sharply just below $T_C$ with the decrease of temperature. This is due to the fact that the spins become randomly oriented to their anisotropy directions with the lowering of temperature or thermal energy. When the sample is cooled under field, the spins try to align to the field direction instead of falling in the anisotropy barrier. This gives rise to the bifurcation between the ZFC and FC curves. We notice that the bifurcation between the ZFC and FC magnetizations is non-monotonic with the grain size. Maximum bifurcation is found in the S850 and S950 samples. This behaviour is clearly observed from the plot of ($M_{FC}$-$M_{ZFC}$) as a function of temperature shown in the inset of Fig. 4. We also notice that the magnitude of ZFC magnetization of the samples also displays non-monotonic change with grain size. This is possibly due to the complicated nature of the phase separation of the samples. This phase separated state is affected non-monotonically by the surface disorder. This also indicates that the intrinsic anisotropy of the samples has a non-monotonic dependence on the grain size in the samples. Obviously, when the disorder effect is very high (for S750) overall magnetization is found to be the lowest as the surface becomes nearly a non-magnetic shell in most of such cases.

Further, we study the linear and non-linear ac magnetic susceptibility of these samples as shown in the Fig. 5. The real part of first order or linear susceptibility ($\chi_1^R$) of all the samples show similar behaviour found in dc magnetization study. In this case also we see a transition at around $T_C$ for all the samples. In the inset of Fig. 5, we have shown the variation of second order ac susceptibility ($\chi_2^R$) as a function of temperature. For the entire series of samples we observe a peak at around $T_C$. We know that $\chi_2$ appears due to the presence of symmetry breaking spontaneous magnetization in a magnetic system [20]. It ($\chi_2$) displays a peak around FM-PM transition because of the onset of spontaneous magnetization. Thus, our results undoubtedly show that the transition at 225 K is really a FM-PM transition.

Next we investigate the electronic-transport properties of our samples and its dependence on grain size. The Fig. 6 displays the resistivity of all the samples as a function of temperature measured with and without the application of magnetic field. We find that the zero field temperature dependent resistivity of the samples exhibit interesting variation with grain size. Firstly, the magnitude of resistivity increases with the decrease of grain size. The overall resistivity increases roughly by one order of magnitude in S750 compared to that of S1150 sample. The S1150 sample displays metallic behaviour ($d\rho/dT>0$) in the measured temperature range. The nature of resistivity and its magnitude for S1150 sample are consistent with those reported previously for bulk polycrystalline $Nd_{0.5}Sr_{0.5}CoO_3$ [15]. Moreover, we observe that all the samples display an anomaly in the resistivity around their corresponding FM-PM transition temperatures. Interestingly, this anomaly takes the form of a metal-insulator transition for S750 showing a peak (inset of Fig. 6(d)) at around 230 K (change of sign of $d\rho/dT$). In an earlier report on $La_{0.5}Sr_{0.5}CoO_3$ system [21], it was found that on reduction of grain size the metallic behaviour is suppressed into an insulating behaviour. However, in our case we observe the metal-insulator transition in a particular sample of nanometric grain size. Furthermore, a significant change in the nature of resistivity with the change of grain size is also observed around the low temperature regime. We find an upturn and shallow minimum in resistivity (at $T_{min}$) for S950 ($T_{min}$= 45 K, Table 3) sample in the low temperature regime. This upturn of resistivity enhances in S850 sample and becomes extremely high sowing up a resistivity minimum at a very high temperature ~ 180 K for S750 sample. The resistivity of the samples show negligible dependence on magnetic field as evident from the resistivity measured at 8 T magnetic field. It is found that only around $T_C$ the magnitude of resistivity is slightly suppressed for all the samples. Such weak field dependence of resistivity is consistent with previous reports on cobaltite systems [4,11]. This also indicates that unlike manganites the processes like spin polarized tunnelling is not significant in our systems.

We now try to analyze and find out the possible mechanism of the temperature dependence of resistivity in our samples. In manganites, the temperature dependent resistivity in the metallic regime has been well described through electron-electron (e-e), electron-magnon or electron-phonon scatterings as reported earlier by a large number of groups [22,23]. However, for cobaltites the electronic transport mechanism is still not explored extensively. Thus, without any pre-assumption we try to fit the ρ vs. T data of S1150 sample in the metallic regime with a general power law,

$$\rho = \rho_0 + \rho_n T^n \text{ -------------------------- (1)}$$

The parameter $\rho_0$ is the residual resistivity. We find a very good fitting as shown in the inset (a) of Fig. 7. Interestingly, the obtained value of $n$ is 2.1. We know that in the metallic regime the inelastic e-e interaction has a $T^2$ dependence [22, 24]. So, we infer from the result of fitting that the dominant interaction in the metallic regime is e-e type. Since n is not exactly equal to 2, there is an indication that other mechanisms might be present. So, next we try an equation of the form,

$$\rho = \rho_0 + \rho_e T^2 + \rho_p T^p \text{ ------------------ (2)}$$

to fit the experimental data assuming that e-e interaction is already present. The last term is supposed to represent any other process like electron-magnon or electron-phonon scatterings. In this case also a very good fitting is obtained as shown in the Fig. 7 (main panel). The obtain value of p is 4.4 (±0.5). We know that the second order electron-magnon scattering goes as $T^{4.5}$. So looking at the value of p we can primarily attribute the third term in right hand side of equation (2) to electron-magnon scattering. However, the error in the value of p indicates that the electron-phonon scattering cannot be completely ruled out which has a $T^5$ dependence in the temperature regime below the Debye temperature. We have applied the equations (1) and (2) in the metallic regime of S950 and S850 samples also. The fits employing equation (2) for S950 and S850 samples are shown in the inset (b) and inset (c) of Fig. 7, respectively. The obtained fitting parameters of all the samples are summarized in the Table 2. Clearly, we find a gradual deviation (increase) of n from 2 with the decrease of grain size when equation (1) is applied. This hints at the fact that scattering processes other than electron-electron type increase when the size of the grains is reduced. This is clarified when equation (2) is employed. We observe an enhancement of the scattering coefficient $\rho_p$ by roughly two orders of magnitude when the grain size is decreased. The exponent p is found to slightly decrease with the decrease of grain size. The e-e scattering coefficient $\rho_e$ which describe the strength of this process obtained from equation (2) is also found to increase with the decrease of grain size.  We also notice that the residual resistivity $\rho_0$ obtained from both the equations (1) and (2), increases with the decrease of grain size. We can attribute this effect (increase of $\rho_0$) to the enhanced grain boundary scattering due to increased crystallographic disorder with the decrease of grain size.

The most fascinating observation in the transport properties is the enhancement of the low temperature minima due to size reduction as mentioned earlier. We attempt to find out the possible origin of this resistivity upturn. Previously, the upturn of resistivity in various materials including manganites was attributed to the phenomena like Kondo effect [25], e-e interaction [26], Coulomb blockade (CB) effect [27] etc. Kondo effect is observed in low resistive diluted magnetic alloys due to the coupling of conduction electron spins with the magnetic impurity moment in the low temperature regime. This gives rise to a low temperature upturn of resistivity which is drastically suppressed on application of magnetic field. As the temperature is increased, various temperature dependent inelastic scattering processes comes to play and as a result a minimum in the resistivity is formed. Assuming Kondo effect the resistivity in the low temperature regime can be written as,

$$\rho = \rho_0 - \rho_s \ln T + \rho_p T^p \text{ -------------- (3)}$$

Here we have included the inelastic scattering term in the form of a general power law as $\rho_p T^p$ which increases with the increase of temperature. Another origin of minima in the resistivity found in different disordered (structural and compositional) materials has been attributed to the elastic e-e interaction. In this case the upturn of resistivity takes place due to quantum interference effect. In such disordered systems, at low temperatures, the mean free path of electrons becomes small and electrons are best described diffusing from site to site by multiple elastic scattering. Previously, the low temperature resistivity minima found in some polycrystalline manganites have been attributed to this e-e interaction effect [28,29]. Including the correction term due to e-e interaction and the inelastic scattering term the resistivity in the low temperature regime can be described as,

$$\rho = \rho_0 - \rho_{ee} T^{1/2} + \rho_p T^p \text{ ------- ------- (4)}$$

On the other hand, the resistivity due to Coulomb blockade effect is empirically given as,

$$\rho = \rho' \exp\left(\sqrt{\frac{\Delta}{T}}\right) \text{ ------------------- (5)}$$

Here $\Delta$ is equivalent to the charging energy of electron. This energy is required as the grains become charged when the electron leaves a grain.

We apply equations (3) to (5) individually and also the different combinations of these equations to fit the low temperature resistivity data of S750 sample. To determine the best fit we plot the difference of experimental data and fitted data with temperature as shown in the inset (a) of Fig. 8. We find that the Coulomb blockade effect (equation (5)) gives extremely bad fit to the experimental data. This indicates that Coulomb blockade effect which is generally expected to be present in granular materials is not responsible for the low temperature upturn of resistivity of S750 sample. We notice that the best fit is obtained on application of equation (4) (Fig. 8 and inset (a)) and the combination of equations (3) and (4). However, when we combine equations (3) and (4), good fit is achieved with negative value of the coefficient $\rho_s$ which is meaningless and unphysical. Furthermore, the observed very small effect of magnetic field on the resistivity minima also indicates that the Kondo effect is very unlikely in our samples. So we conclude that the elastic e-e interaction is the dominant mechanism which gives rise to the low temperature upturn and resistivity minimum in the S750 sample. The resistivity of this sample is of the order of Mott's maximum resistivity (~ 10 mΩ-cm) indicating plausibility of e-e interaction in it. We have also employed the equation (4) on the other samples and observed reasonably good fit to the experimental data (inset (b) of Fig. 8). The obtained parameters are listed in Table 3. From Table 3 it is clear that the e-e interaction coefficient ($\rho_{ee}$) monotonically increases with the decrease of grain size. The coefficient $\rho_{ee}$ is explicitly given by, $\rho_{ee} = 0.0309\sqrt{\kappa_B/\hbar^3 D}\ \rho_0^2 e^2$ [30] where, D is the diffusion constant. So clearly, the residual resistivity $\rho_0$ is also related to $\rho_e$. From Table 3 we observe that $\rho_0$ increases monotonically as $\rho_{ee}$ increases with the decrease of grain size and hence provides a qualitative support to the e-e interaction. We attribute this enhanced e-e interaction with the decrease of grain size to the enhanced disorder in the sample especially in the surface region of the grains. We have already noticed from Fig. 2 (b) that crystallographic disorder is present in the surface region of the lowest grain-size S750 sample. As the size of the grains is reduced the surface to volume ratio is increased magnifying this disorder effect. A similar disorder induced enhancement of the low temperature upturn of resistivity was found in thin films of $SrRuO_3$ [31]. However, the values of $\rho_p$ and p are found to be irregular with grain size. We also notice that $\rho_p$ values are few orders of magnitude lower than $\rho_{ee}$ values, indicating the negligible contribution of these inelastic processes in the low temperature regime. The enhancement of the term $\rho_0$ with decrease of grain size is also consistent with enhanced disorder as $\rho_0$ is also a measure of disorder in the sample [31].

In the paramagnetic insulating regime (> 240 K) of S750 sample, we also try to analyze the resistivity data. We have found that the ρ vs. T data of S750 sample in this regime is best fitted assuming a small polaron hopping model of the form, $\rho = \rho_0 T\exp(E_a/\kappa_B T)$ [32], $E_a$ being the activation energy of the carriers, as shown in the Fig. 9. The calculated $E_a$ is about 24 meV. The variable range hopping model [33] of the form $\rho = \rho_0\exp(T_0/T)^{1/4}$ ($T_0$ is a constant) or even of the form $\rho = \rho_0\exp(T_{ES}/T)^{1/2}$ (Efros-Shklovskii variable range hopping) [34] taking the Coulomb energy into account, do not fit the data satisfactorily as shown in the insets of Fig. 9.

While measuring magnetoresistance [MR (H,T) = $\frac{\rho(H,T) - \rho(0,T)}{\rho(0,T)}$ ] of these samples we observe a peculiar sign reversal of the MR in the low temperature regime. From the MR vs. H curves shown in the Fig. 10 we observe that the S1150 sample displays a very small but positive MR (<1%) at 3 K. The MR of S950 at the same temperature is also positive and small (<0.5%). However a small negative MR is observed for sample S850. This negative value of MR is enhanced upto 4.5% in S750 sample at that temperature. This clearly demonstrates that on reduction of grain size negative MR is enhanced in our samples in the low temperature regime. The positive MR which is a common property of materials, observed in S1150 may be due to the Lorentz force on charge carriers. However, the observed negative MR can be attributed to phase separation effect. It is an extensively discussed issue that in manganites the large negative MR originates partially due to the existence of phase separation and field induced percolation of phases. With the decrease of grain size, possibly the phase separation effect is enhanced in our samples. The negative MR due to this phase separation competes with positive MR and displays highest value in S750. The enhancement of phase separation is, we think, due to the structural disorder which can disturb the spatial extent or the long range correlation of individual phases. Furthermore, we have observed a prominent hysteresis in MR of S750 sample at 3 K (very weak for S850 at 3 K). Such hysteresis was previously attributed to presence of magnetic hysteresis in the corresponding sample [23]. On the other hand, around the FM-PM (225 K) transition all the samples exhibit negative MR as shown in the Fig. 10. This is also visible in the Fig. 6. Such negative MR is generally attributed to the suppression of spin disorder on application of magnetic field in this temperature regime. This can also be partially due to the intrinsic enhancement of the phase separation around the transition temperature of such materials.

## 4. Conclusions

In summary, we have carried out a detail investigation of the electronic-transport properties with additional magnetic characterization of polycrystalline $Nd_{0.5}Sr_{0.5}CoO_3$ with the variation of grain size down to 28 nm. All the samples display a ferromagnetic-paramagnetic transition at around 225 K. We observe a metal-insulator transition in the lowest grain-size sample while the highest grain-size bulk-like sample is metallic up to room temperature. The temperature dependence in the metallic regime of the samples is best described through e-e interaction and electron-magnon or electron phonon interaction. We have also observed a huge enhancement of the low temperature resistivity upturn as the grain size is reduced. This upturn is attributed to elastic e-e interaction generally found in disordered materials. The enhancement of this interaction with the reduction of grain size is attributed to enhanced disorder in (surface effect) in the sample. At very low temperature the observed change of sign of magnetoresistane from positive to negative with the decrease of grain size has been attributed to the enhanced phase separation effect.

## Acknowledgement

One of the authors (T. K. Nath) would like to acknowledge the financial assistance of Department of Science and Technology (DST), New Delhi, India through project no. IR/S2/PU-04/2006.

## References

[1] R. Caciuffo et al., Phys. Rev. B **59**, 1068 (1999).

[2] C. Zobel et al., Phys. Rev. B **66**, 020402(R) (2002).

[3] M. Imada, Rev. Mod. Phys. **70**, 1039 (1998).

[4] N. N. Loshkareva et al., Phys. Rev. B **68**, 024423 (2003).

[5] K. Asai, Phys. Rev. B **40**, 10982 (1989).

[6] Y. Tang et al., Phys. Rev. B **73,** 012409 (2006).

[7] D. N. H. Nam et al., Phys. Rev. B **59**, 4189 (1999).

[8] P. L. Kuhns, Phys. Rev. Lett. B **91**, 127202 (2003).

[9] D. D. Stauffer et al., Phys. Rev. B **70**, 214414 (2004).

[10] A. Ghoshray et al, Phys. Rev. B **69**, 064424 (2004).

[11] R. Mahendiran et al., Phys. Rev. B **68**, 024427 (2003).

[12] M. Paraskevopoulos et al., Phys. Rev. B **63**, 224416 (2001).

[13] K. Yoshii et al., Phys. Rev. B **67**, 094408 (2003).

[14] A. Krimmel et al., Phys. Rev. B **64**, 224404 (2001).

[15] K. Yoshii et al., J. Magn. Magn. Mater. **239**, 85 (2002).

[16] R. K. Pati et al. J. Am. Ceram. Soc., **84,** 2849 (2001).

[17] S. Kundu et al., AIP Conf. Proc. **1349**, 453 (2011).

[18] N. A. Frey Huls et al., Phys. Rev. B 83, 024406 (2011).

[19] R. Ganguly et al., Physica B **271**, 116 (1999); V Sikolenko, J. Phys.: Condens. Matter **21,** 436002 (2009).

[20] A K Pramanik et al., J. Phys.: Condens. Matter **20,** 275207 (2008).

[21] B. Roy et al., Appl. Phys. Lett. **92**, 233101 (2008).

[22] G. Jeffrey Snyder et al., Phys Rev. B **53**, 14434 (1995).

[23] M Ziese, Rep. Prog. Phys. **65,** 143 (2002).

[24] S. Y. Li et al, Phys. Rev. Lett. **93**, 056401 (2004).

[25] J. Kondo, Prog. Theo. Phys. **32**, 37 (1964).

[26] P. A. Lee et al. Rev. Mod. Phys. **57**, 287 (1985);
B. L. Altshuler and A. G. Aronov, Sov. Phys. JETP **50**, 968 (1979).

[27] Ll. Balcells et al., Phys. Rev. B **58**, R14697 (1998).

[28] D. Kumar et al., Phys. Rev. B **65**, 094407(2002).

[29] P. K. Muduli et al. J. Appl. Phys. **105**, 113910 (2009).

[30]E. Rozenberg et al., J. Appl. Phys.**88**, 2578(2000).

[31] G. Herranz et al., Phys. Rev. B **67**, 174423 (2003).

[32] M. Sayer et al., Phys. Rev. B **6**, 4629 (1972).

[33] N. F. Mott and E. A. Davis, *Electronic Processes in Non-Crystalline Materials* (Clarendon, Oxford, 1979), Chap. 2, p. 35.

[34] Y. Noda et al., Phys. Rev B **82**, 205420 (2010).

## Figure captions

**Fig. 1.** (Colour online) The x-ray diffraction pattern of (a) S1150 and (b) S750 samples with the fitted curve and difference plot on Rietveld refinement. The insets display the magnified picture of the main peaks.

**Fig. 2.** (Colour online) (a) TEM image of S750 sample and the distribution of grain size with fitted curve assuming log normal function (inset). (b) High resolution TEM image of grains of S750 sample (marked 1 and 2) showing the lattice structures inside the grains. The arrow indicates the disordered regime around surface. The panels (c), (d) and (e) display the FESEM images of sample S850, S950 and S1150, respectively.

**Fig. 3.** Energy-dispersive x-ray spectroscopic (EDS) curves of (a) S1150 and (b) S750 samples.

**Fig. 4.** (Colour online) Plot of the field-cooled and zero-field-cooled magnetization of all the samples as a function of temperature measured at 100 Oe. The inset shows the plot of $M_{FC}$- $M_{ZFC}$ as a function of temperature of all the samples.

**Fig. 5.** (Colour online) The variation of linear ac magnetic susceptibility ($\chi_1^R$) as a function of temperature for all the samples measured at an ac field of 3 Oe and at fixed frequency of 555.3 Hz. The inset shows the temperature dependent second harmonic of susceptibility ($\chi_2^R$) of the samples.

**Fig. 6.** (Colour online) The measured resistivity of (a) S1150, (b) S950, (c) S850 and (d) S750 samples as a function of temperature at zero field (symbols). The lines are the same measured at 8 T magnetic field. The inset of (d) shows the magnified image around the metal-insulator transition of S750 sample.

**Fig. 7.** (Colour online) The experimental ρ vs. T data (symbols) and fitted curve (line) employing equation (2) for S1150 sample. Inset (a) shows the experimental and fitted curve employing equation (1) for S1150 sample. The insets (b) and (c) exhibit the fitting to the experimental data employing equation (2) for S950 and S850 samples, respectively.

**Fig. 8.** (Colour online) The experimental resistivity data in the low temperature regime with fitted curve on employing equation (4) for S750 sample. The inset (a) shows the difference between the experimental and fitted data employing equations (3) to (5) individually and taking different combinations of them. Inset (b) display the fitting employing equation (4) for S950 sample.

**Fig. 9.** (Colour online) The plot of ln(ρ/T) vs. 1/T (main panel), ln(ρ) vs. $T^{-1/4}$ (inset (a)) and ln(ρ) vs. $T^{-1/2}$ (inset(b)) for S750 sample in the paramagnetic insulating regime. The lines are the fitted curves.

**Fig. 10.** (Colour online) The measured magnetoresistance (MR) of all the samples as a function of magnetic field at different temperatures.

**Table 1**

The lattice parameters, unit cell volume calculated from refinement of the room temperature high resolution x-ray diffraction data and average grain size of the samples.

| Sample | a (Å) | b (Å) | c (Å) | V (Å$^3$) | Average Grain size (nm) |
|---|---|---|---|---|---|
| S1150 | 5.3630(5) | 7.5868(3) | 5.4124(3) | 220.22 | 3700(500) |
| S950 | 5.3663(9) | 7.5866(2) | 5.4120(1) | 220.33 | 610(100) |
| S850 | 5.3677(4) | 7.5912(7) | 5.4133(5) | 220.57 | 310(70) |
| S750 | 5.3776(4) | 7.5883(6) | 5.3956(8) | 220.17 | 28(10) |

**Table 2**

The obtained parameters employing equations (1) and (2) in the metallic regime of S1150, S950 and S850 samples. $R^2$ values of fittings are also given.

| Sample | $\rho = \rho_0 + \rho_e T^2 + \rho_p T^p$ | | | | | $\rho = \rho_0 + \rho_n T^n$ | | | |
|---|---|---|---|---|---|---|---|---|---|
| | $\rho_0$ ($10^{-4}$ Ohm-cm) | $\rho_e$ ($10^{-9}$ Ohm-cm-$K^{-2}$) | $\rho_p$ ($10^{-15}$ Ohm-cm-$K^{-p}$) | p | $R^2$ | $\rho_0$ ($10^{-4}$ Ohm-cm) | $\rho_n$ ($10^{-9}$ Ohm-cm-$K^{-n}$) | n | $R^2$ |
| S1150 | 0.68 | 1.07 | 0.17 | 4.4 | 0.9998 | 0.69 | 0.68 | 2.1 | 0.9998 |
| S950 | 4.93 | 3.36 | 47.13 | 4.0 | 0.9990 | 5.06 | 0.06 | 2.8 | 0.9989 |
| S850 | 9.69 | 4.33 | 25.85 | 4.2 | 0.9988 | 9.90 | 0.03 | 3.0 | 0.9988 |

**Table 3**

The position of resistivity minima ($T_{min}$) and the obtained parameters on employing the formula $\rho = \rho_0 - \rho_{ee}T^{1/2} + \rho_p T^p$ in the low temperature regime of the samples. $R^2$ values of fittings are also given.

| Sample | $T_{min}$ (K) | $\rho_0$ ($10^{-3}$ Ohm-cm) | $\rho_{ee}$ ($10^{-5}$ Ohm-cm-$K^{-1/2}$) | $\rho_p$ (Ohm-cm-$K^{-p}$) | p | $R^2$ |
|---|---|---|---|---|---|---|
| S950 | 45 | 0.53 | 0.76 | $8.52\times10^{-7}$ | 0.7 | 0.9990 |
| S850 | 75 | 1.13 | 1.53 | $2.40\times10^{-15}$ | 5.0 | 0.9714 |
| S750 | 180 | 6.83 | 24.88 | $4.06\times10^{-11}$ | 3.2 | 0.9983 |

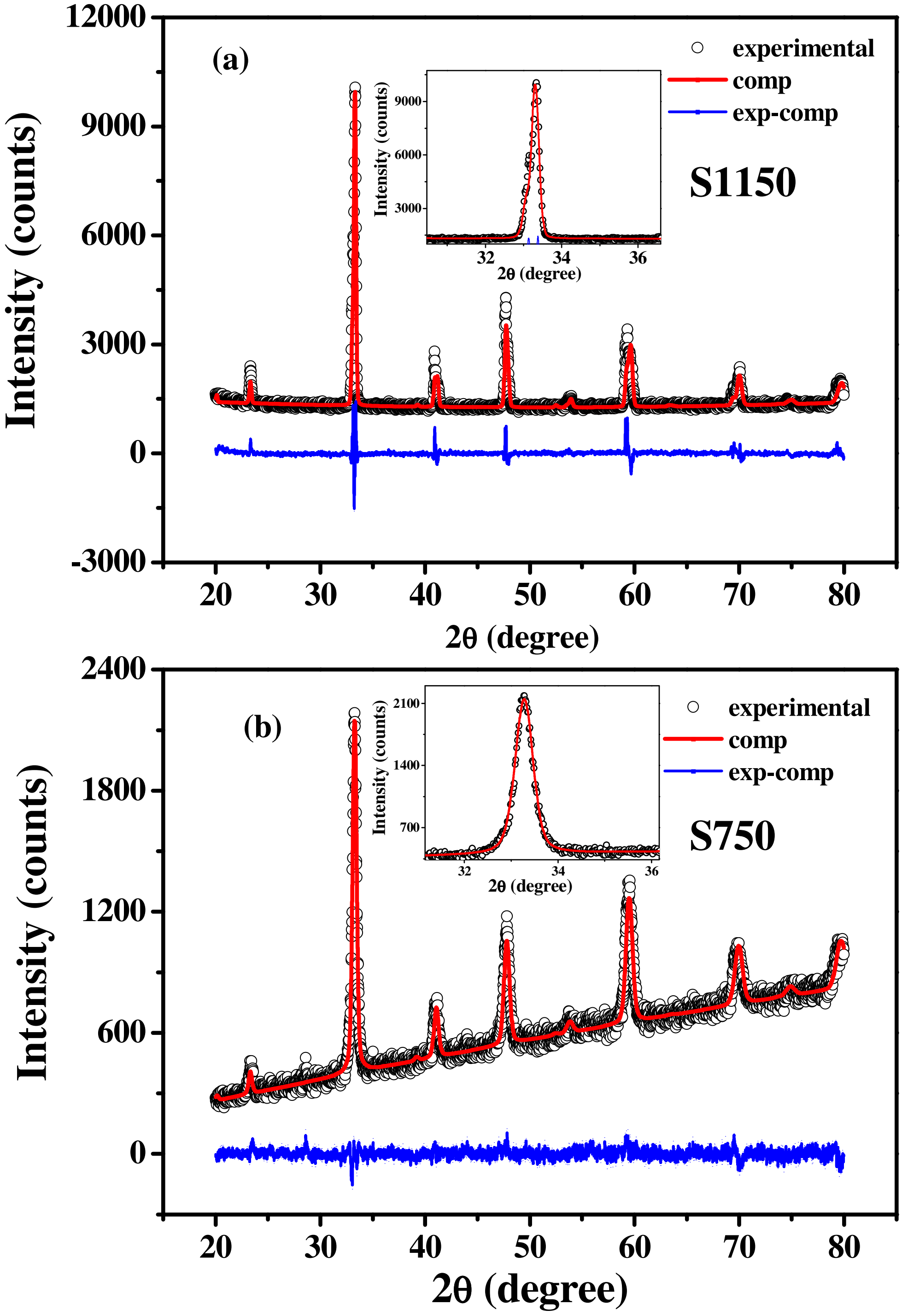

# Fig. 1:S. Kundu et al.

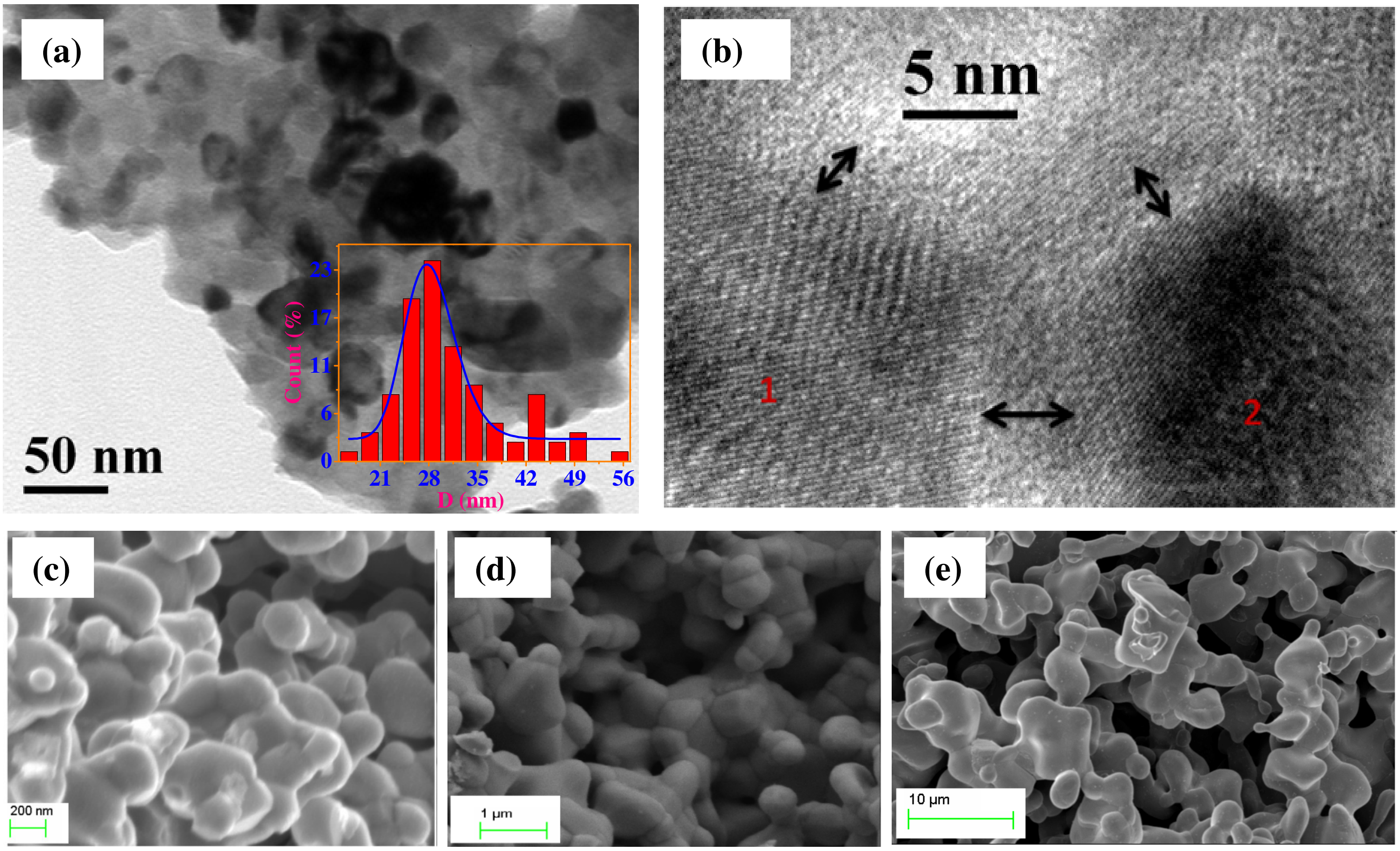

**Fig. 2:S. Kundu et al.**

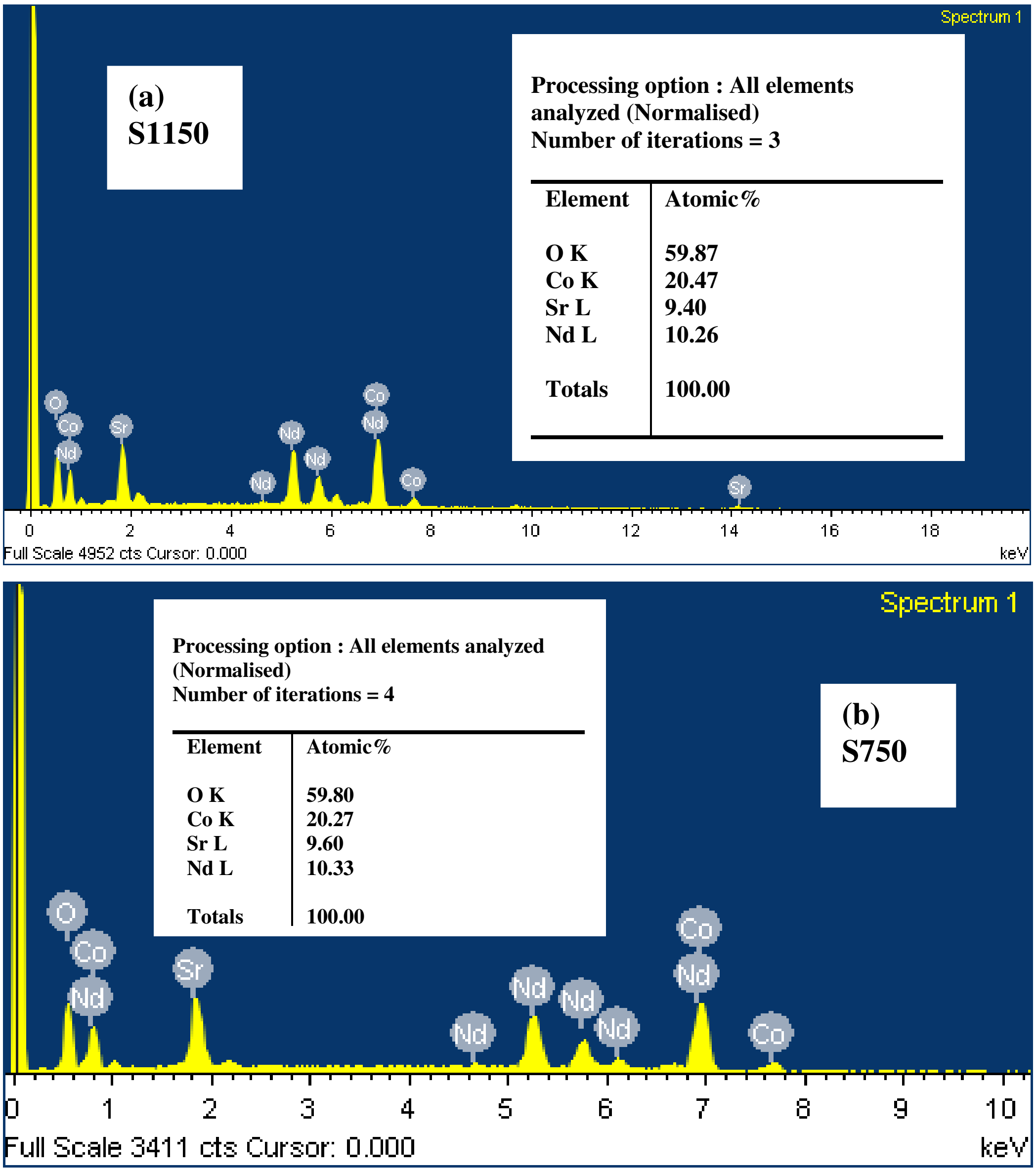

# Fig. 3:S. Kundu et al.

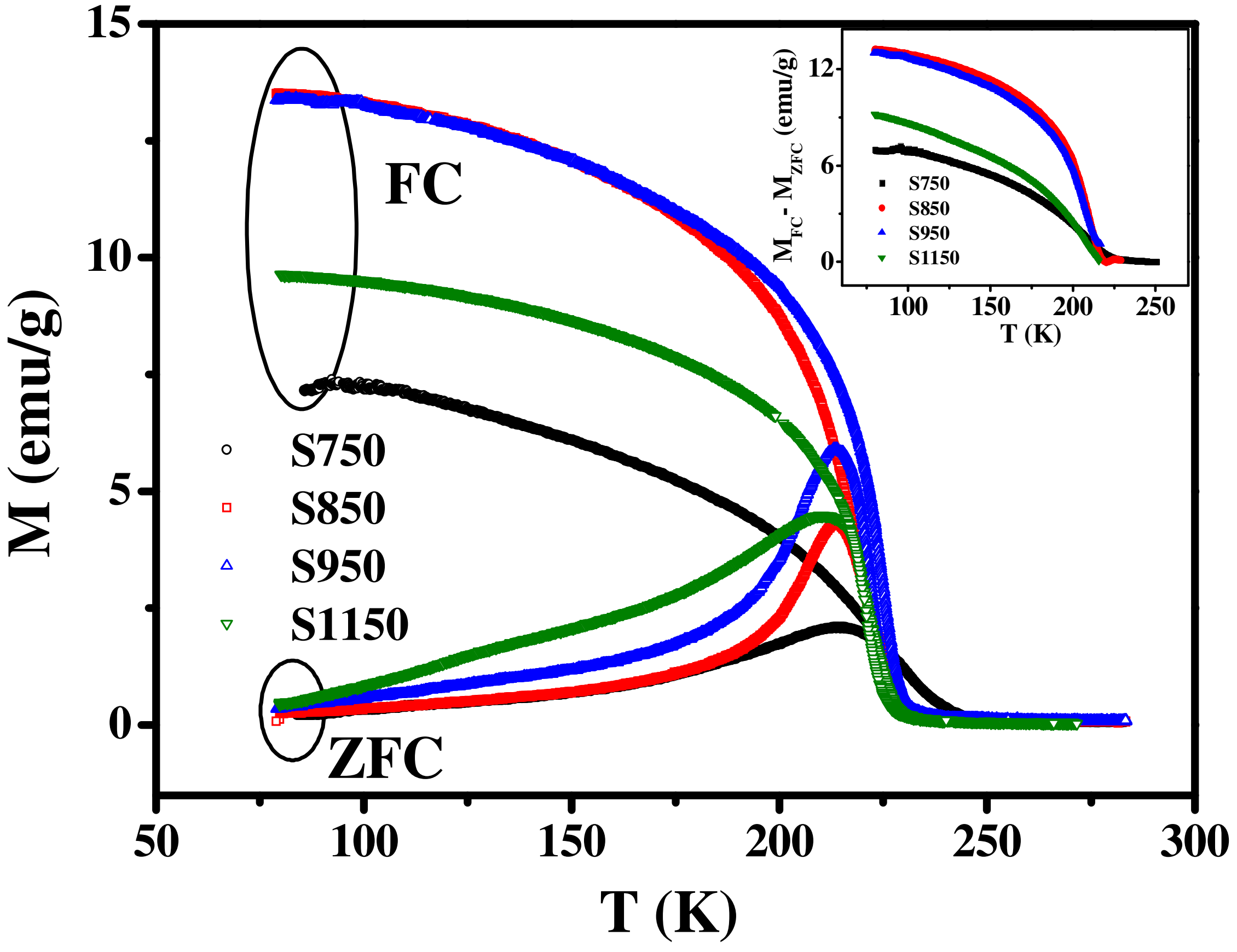

**Fig. 4:S. Kundu et al.**

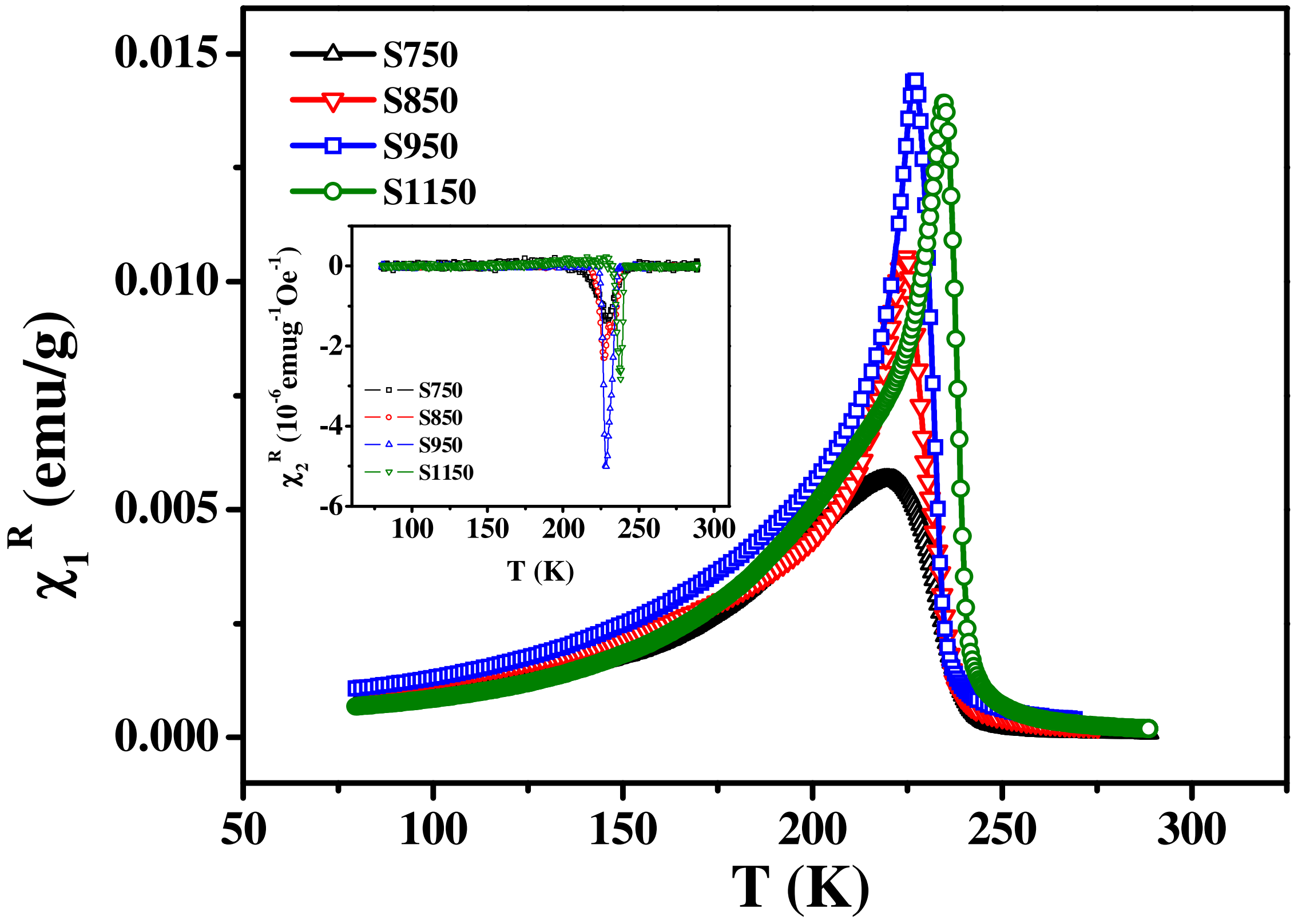

**Fig. 5:S. Kundu et al.**

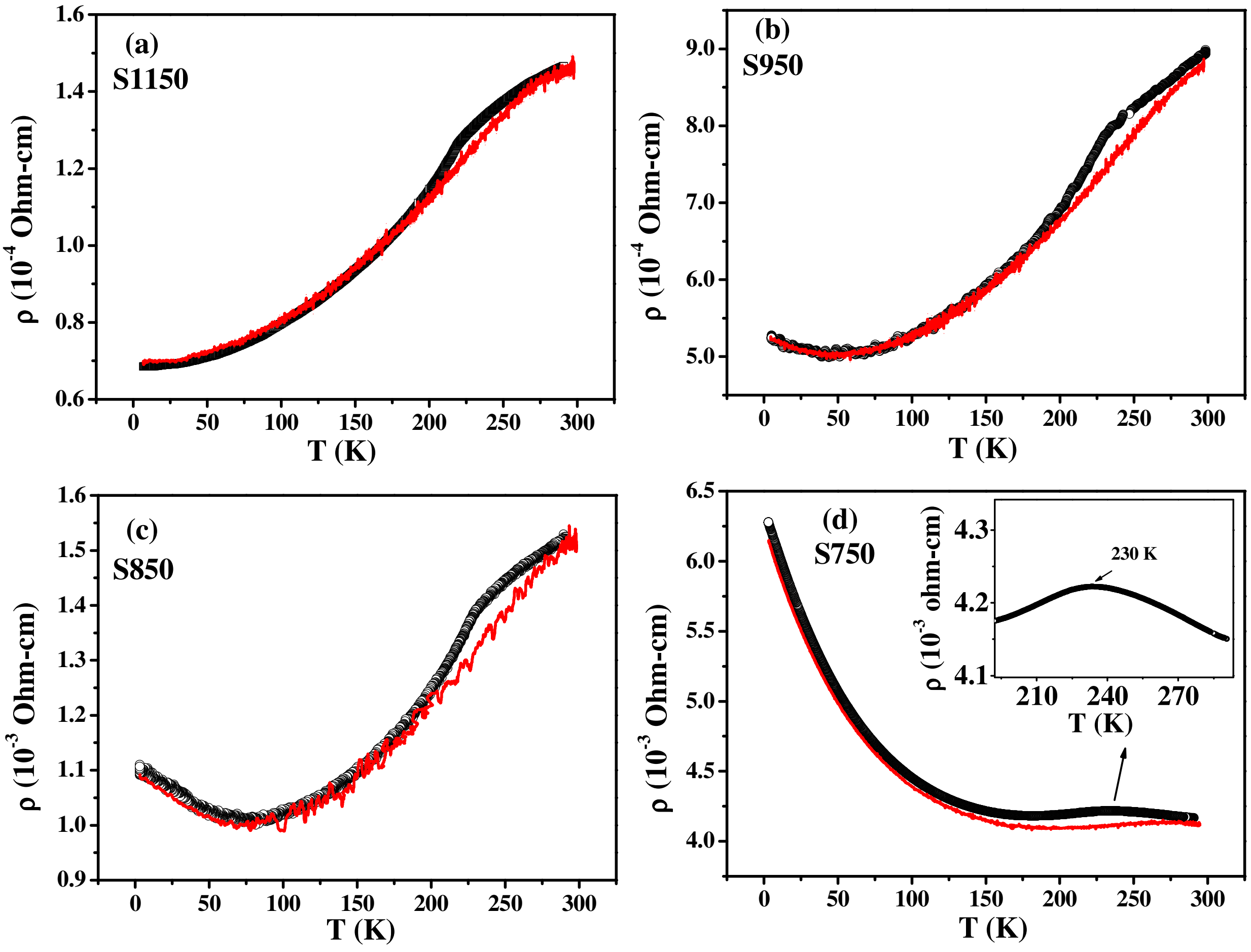

# Fig. 6:S. Kundu et al.

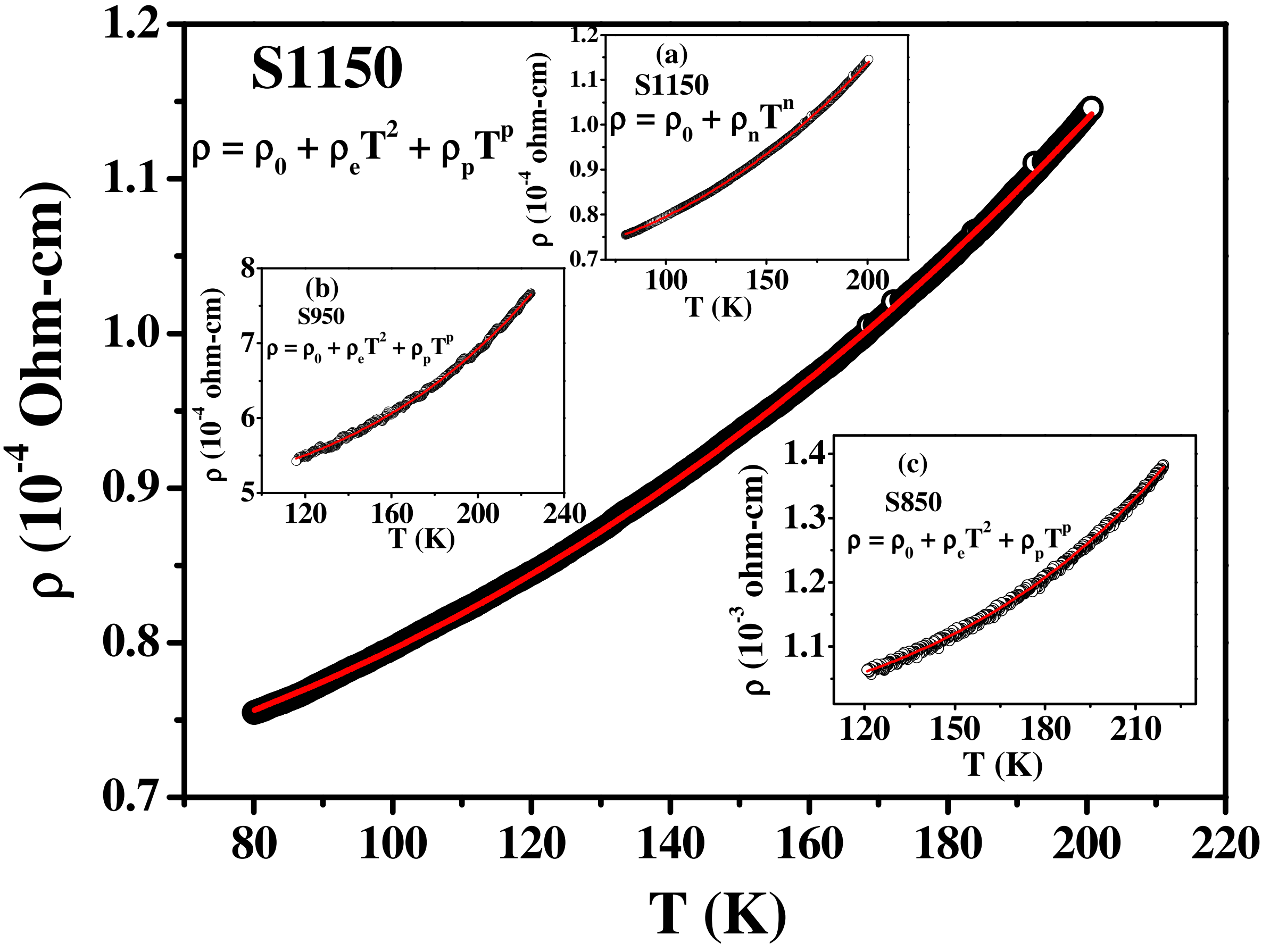

**Fig. 7:S. Kundu et al.**

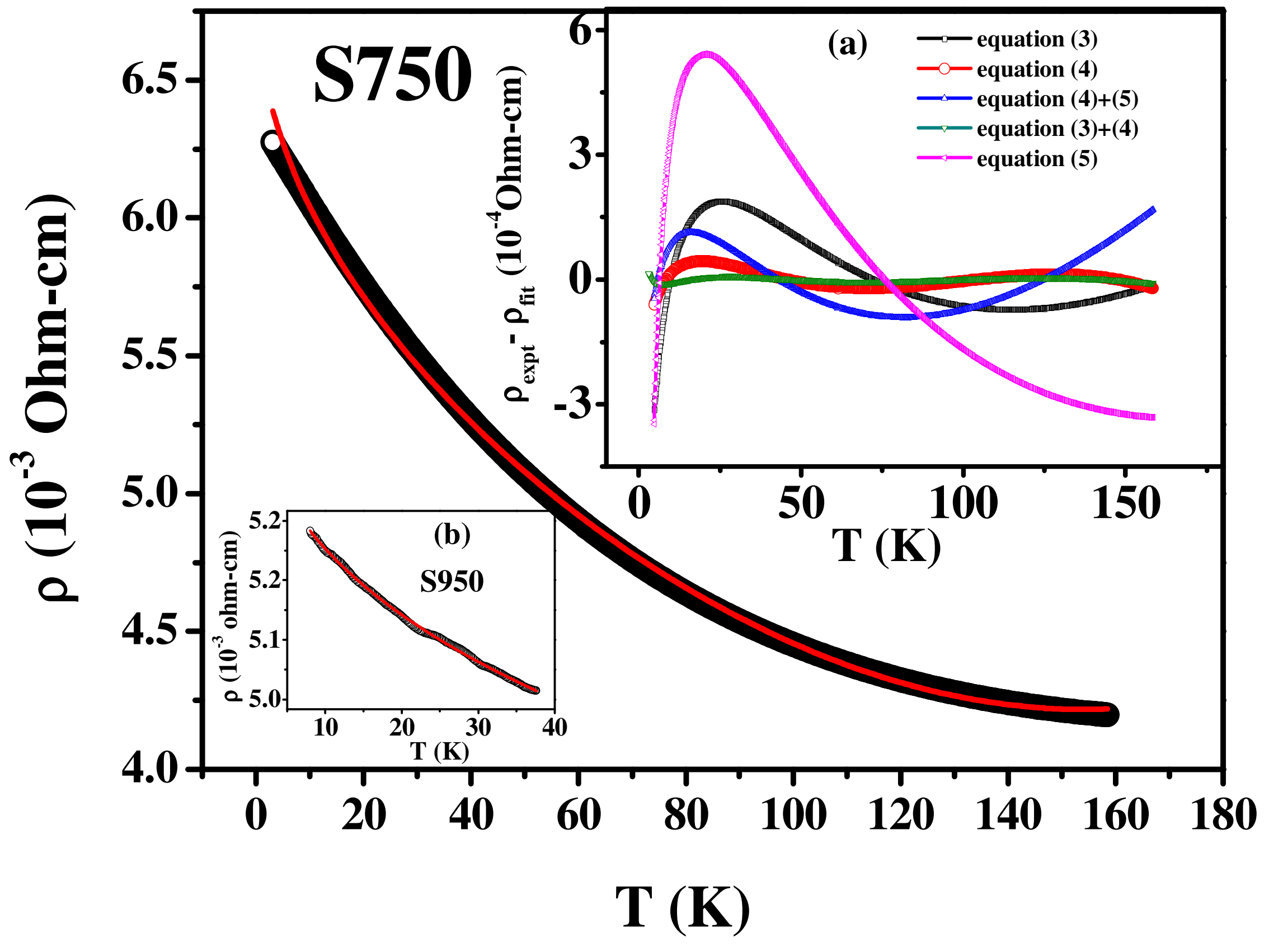

Fig. 8:S. Kundu et al.

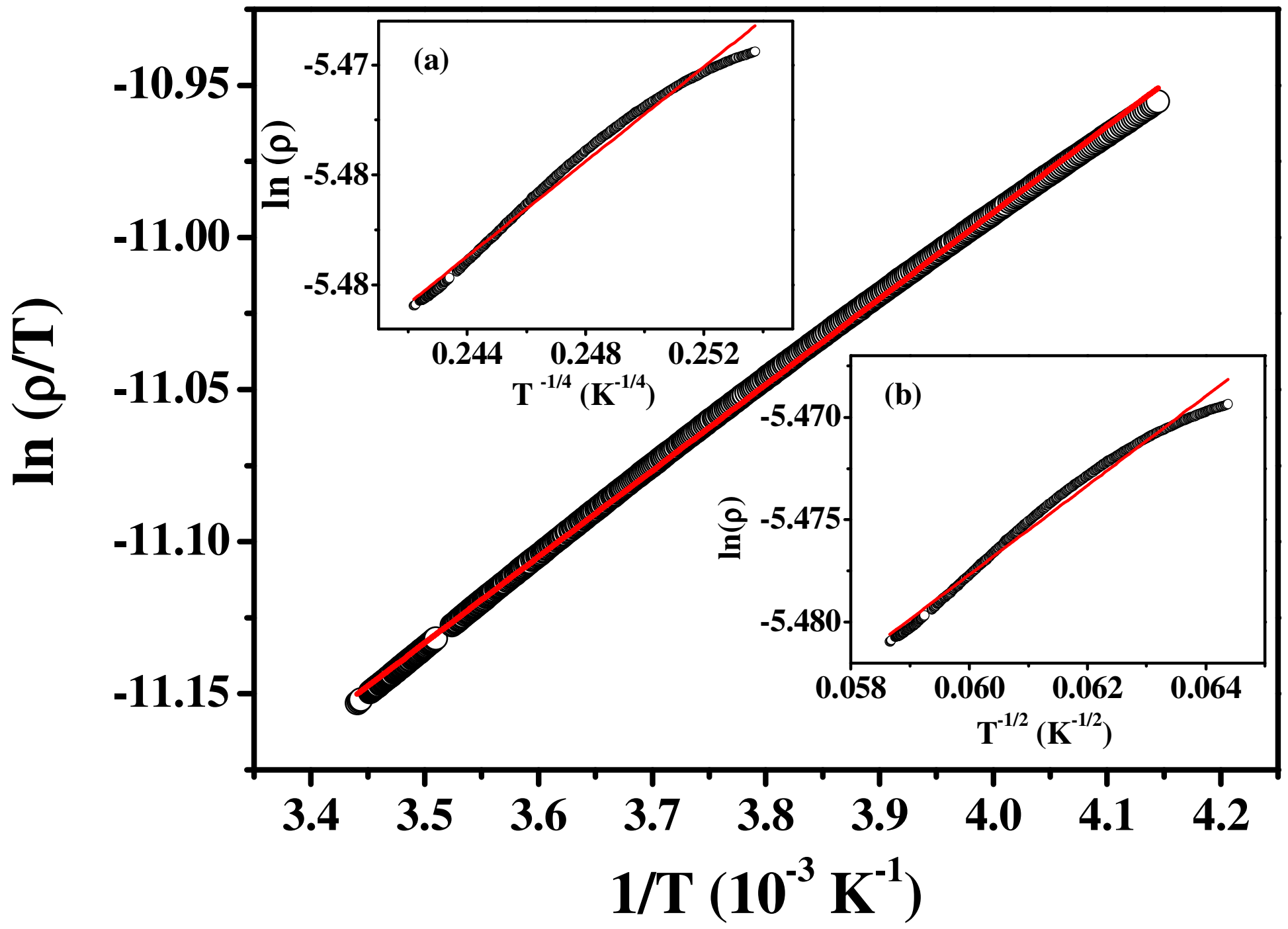

**Fig. 9:S. Kundu et al.**

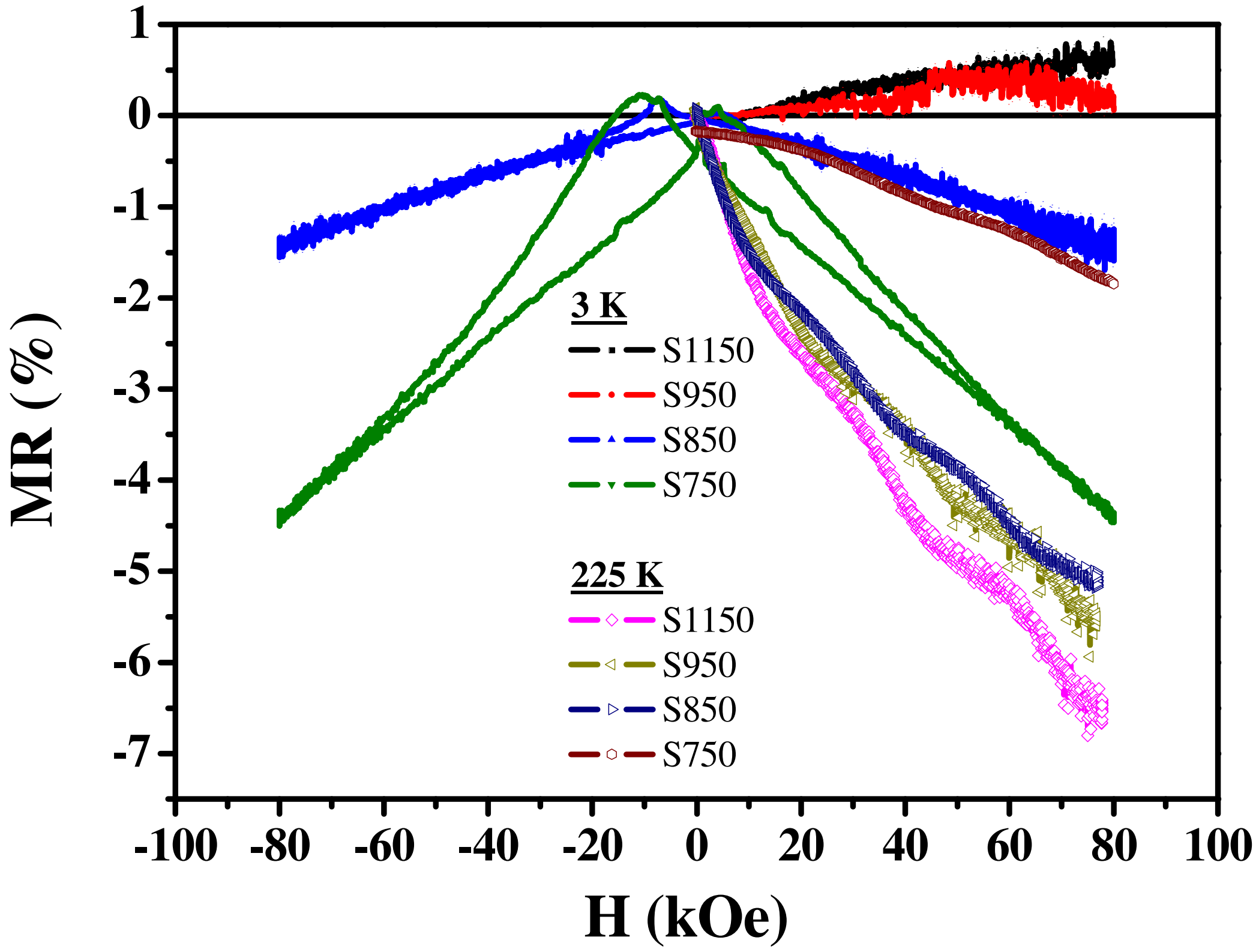

**Fig. 10:S. Kundu et al.**